\documentclass{article}
\usepackage{graphicx} % Required for inserting images
\usepackage{amsmath}
\usepackage{amsfonts}
\usepackage{comment}
\usepackage{amsthm}
\usepackage{amssymb}
\usepackage{url}
\usepackage{hyperref}
\usepackage{xcolor}
\usepackage{natbib}
\newtheorem{theorem}{Theorem}
\newtheorem{proposition}{Proposition}
\newtheorem{lemma}{Lemma}

\newtheorem{corollary}{Corollary}
\newtheorem{observation}{Observation}

\theoremstyle{definition}
\newtheorem{definition}{Definition}
\theoremstyle{definition}
\newtheorem{example}{Example}
\theoremstyle{definition}

\theoremstyle{definition}

\theoremstyle{definition}

\usepackage{algorithm}
\usepackage{algorithmic}
\usepackage{authblk}

\title{Weighted Envy-free Allocation with Subsidy}
\begin{document}

\author[1]{Haris Aziz}

\author[2]{Xin Huang}

\author[2]{ Kei Kimura}
 
\author[2]{Indrajit Saha}

\author[2]{Zhaohong Sun}

\author[1]{Mashbat Suzuki}

\author[2]{Makoto Yokoo}

\affil[1]{UNSW Sydney, Australia}
\affil[2]{ Kyushu University, Japan}

\date{}

\maketitle

\begin{abstract}
We consider the problem of fair allocation of indivisible items with subsidies when agents have weighted entitlements. After highlighting several important differences from the unweighted case, we present several results concerning weighted envy-freeability 
including general characterizations, algorithms for achieving and testing weighted envy-freeability, lower and upper bounds of the amount of subsidies  
for envy-freeable allocations, and algorithms for achieving weighted envy-freeability along with other properties. 
\end{abstract}

\section{Introduction}
\label{sec:intro}
A fundamental problem that often arises in several settings is that of allocating resources in a fair manner. We consider scenarios where agents have valuations over bundles of indivisible items. The goal is to compute allocations of items that are fair. The gold standard for fairness is envy-freeness, which requires that no agent prefers another agent's allocation. For indivisible item allocation, an envy-free outcome may not exist. There are several approaches to achieve envy-freeness. These include randomisation and the use of monetary subsidies. In this paper, we focus on achieving envy-freeness with monetary subsidies.

The literature on envy-free allocation with monetary subsidies / payments / transfers has a long tradition in mathematical economics. For example, the literature on envy-free room-rent division can be viewed as a special case where each agent is supposed to get one item (see, e.g., ~\citep{Klij00a}). More recently, \citet{HaSh19a} studied the problem of finding allocations for which a minimal amount of subsidies will result in envy-freeness. We revisit envy-freeness with subsidies, with one important extension that agents have weighted entitlements. Weighted entitlements, along with weighted envy-freeness, have been considered in many different contexts in fair division
\citep{BEF23a,CISZ21, AMS20a,ACL19a,FGY+19}.

We show that the results under weighted entitlements pose considerable challenges and can often have sharply contrasting results from the unweighted case, i.e., the case of equal entitlements. On the other hand, we also present several results where we generalize some of the celebrated results on envy-freeness with subsidies.

\paragraph{Contributions}

Our first contribution is to show that several celebrated results concerning envy-free allocation with subsidies, do not extend to the weighted case. In particular, the following properties do not hold for the weighted case. 

\begin{enumerate}
\item A welfare maximizing allocation is envy-freeable. 

  \item Given any partition of items, there always exists a way to allocate the bundles in the partition so it is envy-freeable.
    \item For additive valuations, envies can be eliminated by allocating at most one unit of money for each agent.
    \item There always exists an allocation that is both envy-freeable and envy-free up to one item for additive valuations. 
\end{enumerate}
(See Example~\ref{ex:incompatibility-NW-WEFable}, Theorem~\ref{thm:subsidy-lb-additive-identical-items}, and Theorem~\ref{thm:additive-incompatibility-EFable-WWEF1} for counterparts of (1) and (2), (3), and (4), respectively.)

\begin{table}[t]
\caption{We derive upper and lower bounds on worst-case subsidy for each agent in 
weighted envy-freeable allocations under several valuations. We assume that for any agent and any subset of items, the value of the subset for the agent is at most the size of the subset. Here, $n$ represents the number of agents,  $m$ is the number of items, where we assume $n<m$ in this table. The entitlements $w_{max}$ and $w_{min}$ are the maximum and minimum entitlements among the agents, respectively.}
\label{tbl:results}
\begin{center}
\resizebox{\columnwidth}{!}{
\begin{tabular}{ccc}
Valuation &
{Lower bound} & {Upper bound}\\
\hline 

General/ &  $m \frac{w_{max}}{w_{min}}$ 
& $m \frac{w_{max}}{w_{min}}$ \\
Super-modular &[Prop~\ref{prop:subsidy-lb-general}]
& [Prop~\ref{prop:subsidy-ub-general}]\\ 
\vspace{-2mm}& & \\
Matroidal & $\max\{\frac{m}{2}
(\frac{w_{max}}{w_{min}}-1),\frac{w_{max}}{w_{min}}\}$ 
& $m \frac{w_{max}}{w_{min}}$ \\
&[Thm~\ref{thm:subsidy-lb-matroidal},Thm~\ref{thm:subsidy-lb-binaryadditive}]
& [Prop~\ref{prop:subsidy-ub-general}]  \\
\vspace{-2mm}& & \\
Additive & $(n-1) \frac{w_{max}}{w_{min}}$ & $m \frac{w_{max}}{w_{min}}$ \\
&[Thm~\ref{thm:subsidy-lb-additive-identical-items}]
& [Prop~\ref{prop:general-additive}]  \\
\vspace{-2mm}& & \\
Identical additive & 1& 1 \\
&[Thm~\ref{thm:subsidy-lb-identical-additive}]&[Thm~\ref{thm:poly-on-subsidy-identical-additive}]\\
\vspace{-2mm}& & \\
Binary additive & $\frac{w_{max}}{w_{min}}$& $\frac{w_{max}}{w_{min}}$ \\
&[Thm~\ref{thm:subsidy-lb-binaryadditive}]&[Thm~\ref{thm:poly-small-subsidy}]\\
\vspace{-2mm}& & \\
Additive, identical items& $(n-1) \frac{w_{max}}{w_{min}}$& $(n-1) \frac{w_{max}}{w_{min}}+1$\\
&[Thm~\ref{thm:subsidy-lb-additive-identical-items}]&[Thm~\ref{thm:subsidy-ub-additive-identical-items}]\\
\hline
\end{tabular}
}
\end{center}
\end{table}

Nonetheless, we present a generalized characterization of weighted envy-freeness with subsidies by showing its equivalence with two other carefully specified properties. We show that a weighted envy-freeable allocation can be computed and verified in polynomial time. We show further results for the case of super-modular, matroidal, and additive valuations. In particular, we provide upper and lower bounds for worst-case subsidies in %non-wasteful and 
weighted envy-freeable 
allocations under those valuations. The results are summarized in Table~\ref{tbl:results}.

We then present an algorithm that computes envy-free up to one transfer and weighted envy-freeable allocation for two agents.
Finally, we present how to achieve partial fairness when we only have a limited amount of subsidies.

\textbf{Organisations.} This paper is organized as follows. Section~\ref{sec:related_work} reviews related work. Section~\ref{sec:model} defines the mathematical model of the paper. 
Section~\ref{sec:characterization} provides general properties of weighted envy-freeability.
Section~\ref{sec:restricted_Val} gives results for restricted domains such as super-modular and additive valuations.
Section~\ref{sec:app_EF} and \ref{sec:MEF} then provide relaxations of weighted envy-freeness and weighted envy-freeability, respectively.
Finally, Section~\ref{sec:conclusions} concludes the paper.

We omit several proofs, which can be found in the Appendix.

\section{Related Work}
\label{sec:related_work}
\subsection{Envy-free allocation with money}
The literature on envy-free allocation with monetary subsidies / payments / transfers has a long tradition in mathematical economics, e.g.,
\citep{Alkan1991,GMPZ17a,HRS02a,Klij00a,Maskin1987,MPR02a,moulin2004fair,Su99a,%
SunYang2003,Sven83a,Tadenuma1993}.
Most of these works deal with the case where each agent is interested in
at most one item (unit-demand).
With large enough subsidies, an envy-free allocation is guaranteed to exist \citep{Maskin1987} and can be computed in polynomial time \citep{Arag95a,Klij00a}. 

More recently, multi-demand fair division with subsidies has attracted considerable
attention. 
\citet{HaSh19a} showed that
an allocation is envy-freeable with money if and only if the agents cannot increase
social welfare by permuting bundles.
\citet{Brustle2020} showed that 
for additive valuations where the value of each item is at most $1$,
giving at most $1$ to each agent is sufficient to eliminate envies. 
They also showed that for monotone valuations, an envy-free allocation with a subsidy of $2(n-1)$ for each agent exists. 
\citet{Goko2024} developed a truthful mechanism that gives at most $1$ for each agent
when valuations are matroidal, i.e., represented as rank functions of matroids. 
\citet{caragiannis2020computing} studied the computational complexity of approximating the minimum amount of subsidies.

The case where multiple items can be allocated to each agent while the agents pay some amount of money to the mechanism designer, is extensively studied in combinatorial auctions \citep{cramton:2005}. A representative mechanism is the well-known Vickrey-Clarke-Groves (VCG)  mechanism~\citep{clarke,groves:econometrica:1973,vickrey:1961}, 
which is truthful and maximizes social welfare. Envy-freeness is not a central issue in combinatorial auctions, with a notable exception presented by \citet{Papa03b}.

\subsection{Fair allocation with entitlements}

There is a long-standing tradition in fair division to revisit settings and extend them to the case of weighted entitlements. In a classic book by \citet{BrTa96a}, many algorithms and results are extended to the cases of weighted entitlements. This tradition continues in the context of the allocation of indivisible items (see, e.g., \citep{BEF23a,CISZ21, AMS20a,ACL19a,FGY+19}). Recently, \citet{wu2024tree} considered the weighted proportional fairness with subsidies and provided a polynomial time algorithm. % to compute the same.

\section{Model}
\label{sec:model}
We consider the setting in which there is a set $N$ of $n$ agents and a set $M$ of $m$ items. 
We assume each agent $i\in N$ is associated with its
weight $w_i$, where $\sum_i w_i =1$ and 
        $\forall i \in N, w_i>0$ hold.
Let $w_{min} = \min_i w_i$ and 
$w_{max} = \max_i w_i$.

Each agent $i\in N$ has a valuation function $v_i:2^M \rightarrow \mathbb{R}^{+}_{0}$. The function $v_i$ specifies a value $v_i(A)$ for a given bundle $A\subseteq M$. 
 
When $A=\{g\}$, i.e., $A$ contains just one item, 
we often write $v_i(g)$ instead of $v_i(\{g\})$.
We assume the valuation functions are \emph{monotone}, i.e., for each $i\in N$ and $A \subseteq B \subseteq M$,  $v_i(A)\le v_i(B)$.
When we examine the subsidy bounds, we assume the valuation of each agent is bounded, i.e.,  
for any $i\in N$, $A \subseteq M$, $v_i(A)\leq |A|$ holds.

The valuation function of an agent $i$ is \emph{super-modular} if for each $i\in N$, and $A,B\subseteq M$,  $v_i(A\cup B)\geq v_i(A)+v_i(B)-v_i(A\cap B)$. 
The valuation function of an agent $i$ is \emph{additive} if for each $i\in N$, and $A,B\subseteq M$ such that $A\cap B=\emptyset$, the following holds: $v_i(A\cup B)= v_i(A)+v_i(B)$.
The valuation function of agent $i$ is \emph{binary additive} if
it is additive and for each $i \in N$ and $g\in M$, 
$v(g)\in \{0, 1\}$ holds.

An \emph{allocation} $X=(X_1,\ldots, X_n)$ is a partitioning of the items into $n$ bundles where $X_i$ is the bundle allocated to agent $i$.
We assume allocation $X$ must be \emph{complete}, i.e., 
$\bigcup_{i \in N} X_i = M$ holds; each item must be allocated to some agent.
For an allocation $X$, the classical weighted social welfare $SW(X)$ is $\sum_{i\in N}w_i \cdot v_i(X_i)$.

An \emph{outcome} is a pair consisting of the allocation and the subsidies
received by the agents. Formally, an outcome is a pair $(X,p)$ where $X=(X_1,\ldots ,X_n)$ is the allocation that specifies bundle $X_i\subseteq M$ for agent $i$ and $p \in (\mathbb{R}^+_0)^n$ specifies the subsidy $p_i$ received by agent $i$.

An agent $i$'s \emph{utility} for a bundle-subsidy pair $(X_j,p_j)$ is $v_i(X_j)+p_j$. In other words, we assume quasi-linear utilities.
An outcome $(X,p)$  is \emph{envy-free} if for all $i,j\in N$, it holds that $v_i(X_i)+p_i\geq v_i(X_j)+p_j$. 
%An outcome $(X,p)$  is \emph{equitable} if for all $i,j\in N$, $u_i(X_i,p_i)=u_j(X_j,p_j)$.
An allocation $X$ is \emph{envy-freeable} if there exists a subsidy vector $p$ such that $(X,p)$ is envy-free. 

A mechanism is a function from the profile of 
declared agents' valuation functions to an outcome. 
We say a mechanism is truthful if no agent can obtain a 
strictly better outcome by misreporting its valuation function.

\begin{definition}[Weighted envy-freeability]
An outcome $(X,p)$ is \emph{weighted envy-free} 
if for all $i, j \in N$:
$$\frac{1}{w_i}(v_i(X_i)+p_i)\geq \frac{1}{w_j}(v_i(X_j)+p_j).$$

    An allocation $X$ is \emph{weighted envy-freeable} if there are payments $(p_1,\ldots, p_n)$ for agents such that 
    $(X,p)$ is weighted envy-free. 
\end{definition}

\begin{example}
Assume a family tries to divide inheritance. Agent 1 is the spouse, whose weight is $1/2$. 
Agents 2 and 3 are children, whose weights are $1/4$. 
There are two items: $g_1$ is a house, and $g_2$ is a car. 
Some money is also left, but the testament says the money can be divided among agents
only to make the outcome weighted envy-free; the remaining amount should be donated to charity. Assume $v_1(g_1)=100, v_2(g_1)=70, v_3(g_1)=0$, and 
$v_1(g_2)=40, v_2(g_2)=60, v_3(g_2)=0$.

Intuitively, between two agents $i$ and $j$, $w_i/w_j$ represents the relative importance of agent $i$ against $j$. Here, 
the spouse is twice more important than a child, and should get twice more inheritance. 
Here, agent $3$ is not interested in these items, but still cares about the payments. 
There are two weighted envy-freeable allocations: $(\{g_1, g_2\}, \emptyset, \emptyset)$, i.e., 
allocating both items to agent 1, and 
$(\{g_1\}, \{g_2\}, \emptyset)$, i.e., agent 1 obtains $g_1$, while agent 2 obtains $g_2$.
For the first allocation, we need to pay $65$ to agents 2 and 3. 
For the second allocation, no subsidy is needed; the allocation is weighted envy-free.
\end{example}

Let us introduce several properties related to agents' welfare.
\begin{definition}[Pareto efficiency]
We say allocation $X$ dominates another allocation $X'$ if 
$\forall i\in N, v_i(X_i)\geq v_i(X'_i)$ and 
$\exists j\in N$, $v_j(X_j) > v_j(X'_j)$ hold. 
We say $X$ is Pareto efficient %(PE) --- seems we don't use PE anywhere
if it is not dominated by any other allocation. 
\end{definition}
\begin{definition}[Non-wastefulness]
We say allocation $X$ is \emph{non-wasteful} if $\forall i \in N$, $\forall g \in X_i$, if 
$v_i(X_i)= v_i(X_i \setminus \{g\})$ holds, then $v_j(X_j \cup  \{g\})=v_j(X_j)$ for all $j\neq i$. 
\end{definition}
In other words, no item can be transferred from one agent to another, and the transfer results in a Pareto improvement. 

\begin{definition}[Weighted welfare maximizing allocation]
We say allocation $X$ maximizes the weighted social welfare if 
for any allocation $X'$, 
$SW(X) \geq SW(X')$ holds. 
\end{definition}

It will be shown that weighted envy-freeability and non-wastefulness are incompatible in general (Example~\ref{ex:incompatibility-NW-WEFable}).
For this reason, let us introduce yet another very weak efficiency property. 
\begin{definition}[Non-zero social welfare]
We say allocation $X$ satisfies non-zero social welfare property if
$SW(X)=0$, then for any other allocation $X'$, $SW(X')=0$ holds. 
\end{definition}
This property means choosing $X$ s.t. $SW(X)=0$ is allowed only when social welfare is 0 for any allocation. 

A weighted welfare maximizing allocation is Pareto efficient, but not vice versa.
Pareto efficiency implies non-wastefulness, as well as non-zero social welfare, but 
not vice versa. 
Non-wastefulness and non-zero social welfare are independent. 

\section{A General Characterization and its Implications}
\label{sec:characterization}
We first give a general characterization of weighted envy-freeable allocations. For the characterization, we generalize a couple of previously studied mathematical objects to the weighted case. 

\begin{definition}[Weighted reassignment-stability]
We say that an allocation $X$ is \emph{weighted reassignment-stable} if 
\begin{align}\label{eq:WRS}
\sum_{i\in N} \frac{v_i(X_i)}{w_i}\geq \sum_{i\in N} \frac{v_i(X_{\pi(i)})}{w_{\pi(i)}}
\end{align}
for all permutations $\pi$ of $N$. 
\end{definition}

\begin{definition}[Weighted envy-graph]
For any given allocation $X$, the corresponding \emph{weighted envy-graph} 
is a complete directed graph with vertex set $N$. 
For any pair of agents $i,j\in N$, $\ell(i,j)$ is the length of edge  $(i,j)$, 
which presents the fact that agent $i$ has envy toward agent $j$ under the allocation $X$:  
\[\ell(i,j) \ =\  \frac{1}{w_j}v_i(X_j)-\frac{1}{w_i}v_i(X_i).\] For any path or cycle $C$ in the graph, $\ell(C)$ is the length of the $C$, which is the sum of lengths of edges along $C$.
\end{definition}

\begin{theorem}
	\label{th:ef}
	The following conditions are equivalent for a  given allocation:
	\begin{enumerate}
		\item the allocation is weighed envy-freeable; 
		\item the allocation is weighted reassignment-stable;
		\item for the allocation, there is no positive length cycle in the corresponding weighted envy-graph.
 
	\end{enumerate}
	\end{theorem}
\begin{proof}
$(1) \Rightarrow (2)$. Suppose the allocation $X$ is weighted envy-freeable. Then, there exists a payment vector $p$ such that for all agents $i,j$ 
$\frac{1}{w_i}(v_i(X_i)+p_i)\geq \frac{1}{w_j}(v_i(X_j)+p_j)$. Equivalently, 
$\frac{1}{w_j}v_i(X_j)- \frac{1}{w_i}v_i(X_i)\leq \frac{1}{w_i}p_i - \frac{1}{w_j}p_j$.
Consider any permutation $\pi$ of $N$. Then,
$$\sum_{i\in N} \left(\frac{1}{w_{\pi(i)}}v_i(X_{\pi(i)})- \frac{1}{w_i}v_i(X_i)\right)\leq \sum_{i\in N}\bigg(\frac{1}{w_i}p_i - \frac{1}{w_{\pi(i)}}p_{\pi(i)}\bigg)=0.$$ The last entry is zero as all the weighted payments are considered twice, and they cancel out each other. 
Hence the allocation $X$ is weighted reassignment-stable.

$(2) \Rightarrow (3)$.
Suppose some allocation $X$ has a corresponding weighted envy-graph with a cycle $C$ of strictly positive length. Then consider a permutation $\pi$ such that $\pi(i)=i$ if $i\notin C$ and  if $i\in C$, then 
$\pi(i)$ is the agent that $i$ points to in $C$. In that case 
 \[\sum_{i\in N} \frac{v_i(X_i)}{w_i}< \sum_{i\in N} \frac{v_i(X_{\pi(i)})}{w_{\pi(i)}},\] which means that $X$ is not weighted reassignment-stable.

$(3) \Rightarrow (1)$.
Suppose (3) holds. Let $\ell_i$ be the maximum length of any path in the weighted envy-graph that starts from $i$, which is well-defined since there is no positive length cycle. Let each agent $i$'s payment be $p_i=\ell_i\cdot w_i$.
Then 
$$p_i/w_i=\ell_i\geq \ell(i,j)+\ell_j=\frac{1}{w_j}v_i(X_j)-\frac{1}{w_i}v_i(X_i) + p_j/w_j.$$
This implies that $(X,p)$ is weighted envy-free, and hence $X$ is weighted envy-freeable. 
\end{proof}

Our characterization suggests a simple method for 
testing whether a given allocation is weighted envy-freeable.

\begin{proposition}
Given an allocation $X$, it can be checked in polynomial time whether $X$ is weighted envy-freeable.
\end{proposition}
\begin{proof}
    Checking whether $X$ is weighted envy-freeable is equivalent to checking if $X$ is weighted reassignment-stable. Consider a complete bipartite graph $G$ that has vertex set $N$ on one side and a copy of $N$ on the other side and edge set $N \times N$, and the cost of each edge $\{i, j\} \in N \times N$ is defined as $v_i(X_j)/w_j$. Then, it follows the definition of weighted reassignment-stable that $X$ is weighted reassignment-stable if and only if its induced matching in $G$ is a maximum weight perfect matching, which can be checked in 
    polynomial time~\citep{edmonds1972theoretical}. 
\end{proof}

The following theorem is similar to \citep[Theorem 2]{HaSh19a}, which states the minimum subsidy required when given a weighted envy-freeable allocation.
\begin{theorem}\label{thm:minsubsidy}
    Given a weighted envy-freeable allocation, the minimum subsidy for each agent is the length of the longest path in a weighted envy-graph times her weight.
\end{theorem}
\begin{proof}
    Let $p_i$ be the subsidy for the agent $i$. By the definition of weighted envy-freeness, we have $\frac{v_i(X_i)+p_i}{w_i}\ge\frac{v_i(X_j)+p_j}{w_j}$ for any $i$ and $j$. 
    
    Let $U_i$ be the longest path from $i$ and $\ell_i$ be the length of the path.  Sum up $\sum_{(i',j')\in U_i}\frac{v_{i'}(X_{i'})+p_{i'}}{w_{i'}}\ge
    \sum_{(i',j')\in U_i}\frac{v_{i'}(X_{j'})+p_{j'}}{w_{j'}}$. Then we have $\frac{p_i}{w_i}-\frac{p_k}{w_k}\ge \ell_i$ where $k$ is the last agent in the path $U_i$.
    As the subsidy is non-negative, we have $p_i\ge w_i\cdot \ell_i$.

    On the other hand, if we let $p_i=w_i\cdot \ell_i$,
    then it is weighted envy-freeable. Let $\ell_i$ (and $\ell_j$) 
    be the length of the longest path from $i$ (and $j$). 
    We have $\ell_i\ge \ell(i,j)+\ell_j$. Therefore, weighted envy-freeable is implied by $\frac{v_i(X_i)+p_i}{w_i}-\frac{v_i(X_j)+p_j}{w_j}=\ell_i-\ell(i,j)-\ell_j\ge0$.
\end{proof}

The following lemma is useful for showing weighed envy-freeability and the subsidy bounds. 
\begin{lemma}
\label{lem:max-value}
For allocation $X$, if for all $i, j \in N$, 
$v_i(X_i)\geq v_j(X_i)$ holds, then $X$ is weighted envy-freeable, 
the length of a path from agent $i$ to $j$ is bounded by 
$\frac{v_j(X_j)}{w_j}-\frac{v_i(X_i)}{w_i}$, 
and the maximum subsidy 
for each agent is bounded by $m\frac{w_{max}}{w_{min}}$.
\end{lemma}
\begin{proof}
For the sake of contradiction, assume $X$ is not weighted envy-freeable. 
Thus, it does not satisfy 
reassignment-stability, i.e., there exists permutation $\pi$
where $$\sum_{i\in N} v_i(X_i)/w_i < \sum_{i\in N} v_i(X_{\pi(i)})/w_{\pi(i)}~\text{holds}.$$  Let $\pi^{-1}$ denote the inverse function of $\pi$. 
Then, $\sum_{i\in N} (v_i(X_i)-v_{\pi^{-1}(i)}(X_i))/w_i < 0$ must hold. However, this contradicts the fact that 
for each $i,j \in N$, $v_i(X_i)\geq v_j(X_i)$ holds. 

Next, consider path $P$ from $i$ to $j$ in the weighted envy-graph.
We show $\ell(P) \le \frac{v_j(X_j)}{w_j}-\frac{v_i(X_i)}{w_i}$ holds. 
        As for any $h$ and $k$, we have $v_h(X_k)\le v_k(X_k)$.
        Using this, we have 
            $\ell(P)=\sum_{(h,k)\in P}\ell(h,k)
            =\sum_{(h,k)\in P} \frac{v_h(X_k)}{w_k}-\frac{v_h(X_h)}{w_h}
            \le \sum_{(h,k)\in P}\frac{v_k(X_k)}{w_k}-\frac{v_h(X_h)}{w_h}
            =\frac{v_j(X_j)}{w_j}-\frac{v_i(X_i)}{w_i}$.
Since $\frac{v_j(X_j)}{w_j}-\frac{v_i(X_i)}{w_i}\leq \frac{m}{w_{min}}$, 
the subsidy for $i$ is bounded by $m\frac{w_{max}}{w_{min}}$.
\end{proof}

Next, we make some observations that highlight why several aspects do not extend when generalizing to the case of weighted entitlements.

\begin{observation}
In the unweighted case, any welfare maximizing allocation is envy-freeable.
In the weighted case, there exists an instance where
% Pareto efficient allocation, as well as 
any weighted welfare maximizing allocation, as well as any non-wasteful allocation, 
is not weighted envy-freeable.
\end{observation}
\begin{observation}
In the unweighted case, for any partition of $M$ into $n$ bundles, 
we can always find an envy-freeable allocation based on the partition. 
In the weighted case, there exists an instance and a partition of the items into bundles such that no assignment of the bundles to the agents results in a weighted envy-freeable allocation.
\end{observation}

These two observations are derived from the following example.

\begin{example}
\label{ex:incompatibility-NW-WEFable}
Consider the case with two agents $1, 2$, with weights $3/4, 1/4$, respectively. 
There are two identical items. 
Agent 1 values one item as 90, while agent 2 values one item as 30. 
The marginal utility for the second item is 0 (these items are substitutes). 
The weighted social welfare is maximized
by allocating one item for each agent. Also, this is the only non-wasteful allocation.

This allocation does not satisfy weighted reassignment-stable: 
$\frac{90}{w_1} + \frac{30}{w_2} = 240 < 
\frac{90}{w_2} + \frac{30}{w_1} = 400.$
Also, consider a partition where each bundle contains one item, 
there is no weighted envy-freeable allocation based on this partition. 
\end{example}

\begin{proposition}\label{prop:subsidy-ub-general}
There always exists a weighted envy-freeable and non-zero social welfare allocation. 
Furthermore, the maximum subsidy for each agent 
is at most $m \frac{w_{max}}{w_{min}}$. 
\end{proposition}
\begin{proof}
Allocate all items to agent $i^*$, where $v_{i^*}(M)$ is the largest. 
It is clear that this allocation satisfies non-zero social welfare property. 
Also, $v_i(X_i) \geq v_j(X_i)$ holds for any $i, j \in N$. 
Thus, by Lemma~\ref{lem:max-value}, the allocation is weighted envy-freeable and
the subsidy for each agent is bounded by $m\frac{w_{max}}{w_{min}}$.

\end{proof}

\begin{proposition}\label{prop:subsidy-lb-general}
%Assume for any $B \subseteq M$ and $i\in N$, $v_i(B)\leq |B|$ holds. 
%Then, 
The subsidy for each agent can be $m \frac{w_{max}}{w_{min}}$ and the total amount of subsidies can be $(n-1)m \frac{w_{max}}{w_{min}}$
for a non-zero social welfare allocation.
\end{proposition}
\begin{proof}
Let us assume all agent has an all-or-nothing valuation for $M$.     
$w_1=w_{min}$ and $w_2=w_3 = \ldots w_n= w_{max}$.
$v_1(M)=m$ and $v_i(M)=m-\epsilon$ for $i\neq 1$. 
Then, the only weighted envy-freeable and non-zero social welfare allocation is allocating $M$ to agent 1. 
We need to pay $(m-\epsilon) \frac{w_{max}}{w_{min}}$ for the rest. 
\end{proof}
Note that an all-or-nothing valuation is one instance of a super-modular valuation. 
Thus, this bound also holds for a super-modular valuation.

\begin{observation}
Negative results from the unweighted setting carry over to the weighted case, 
since the unweighted case is equivalent to the weighted case where each weight $w_i = 1/n$. For example, it is NP-hard to compute the minimum
subsidy required to achieve (weighted) envy-freeness even in the binary additive case, 
assuming the allocation is non-wasteful
(by \citep[Corollary 1]{HaSh19a}).
\end{observation}

\section{Restricted Domains}
\label{sec:restricted_Val}
In this section, we present weighted envy-freeable allocations and the subsidy bounds for special classes of valuations. 

\subsection{Super-modular valuation}
The condition of super-modularity can be re-written as follows: 
for any $X, Y, Z \subseteq M$, where $X \subseteq Y$, 
$v_i(X\cup Z) - v_i(X) \leq v_i(Y \cup Z) - v_i(Y)$, by 
choosing $X=A\cap B$, $Y=B$, and $Z=A\setminus B$.
This condition means the marginal value for adding $Z$ weakly increases 
when the original bundle becomes larger.

\begin{definition}[(Unweighted) social welfare maximization]
For a subset of agents $S\subseteq N$ and 
a subset of items $A \subseteq M$, 
let ${\mathcal X}^{S,A}$ denote all possible allocations of $A$ among $S$. 
We call $X^{S,A}\in {\mathcal X}^{S,A}$ is a (unweighted) social welfare maximizing 
allocation with respect to $S$ and $A$ if $\sum_{i \in S} v_i(X^{S,A}_i) \geq 
\sum_{i \in S} v_i(X_i)$ holds for any $X \in {\mathcal X}^{S,A}$. 
If $S = N$ and $A=M$, we omit ``with respect to $N$ and $M$'' and just say a (unweighted) social welfare maximizing 
allocation.
\end{definition}
For allocation $X$, let $V(X)$ denote $\sum_{i\in N} v_i(X_i)$. 
Any unweighted social welfare maximizing allocation is Pareto efficient, 
but not vice versa.

\begin{theorem}
When valuations are super-modular, an unweighted social welfare maximizing allocation is weighted envy-freeable.
\end{theorem}
\begin{proof}
We show that for an unweighted social welfare maximizing allocation $X$, $v_i(X_i) \geq v_j(X_i)$ holds for any $i, j \in N$. For the sake of contradiction, assume $v_i(X_i) < v_j(X_i)$ holds. In this case, we can construct another allocation $X'$, 
where for all $k \neq i, j$, $X'_k=X_k$, 
$X'_i =\emptyset$, $X'_j = X_j \cup X_i$. In other words, we reassign $X_i$ from $i$ to $j$. By the definition of super-modularity, we have $v_j(X_j') = v_j(X_i \cup X_j) \geq v_j(X_i) + v_j(X_j)$. Therefore, the total social welfare under allocation $X'$ is
\[
V(X') = v_j(X_j') + \sum_{k\neq i,j} v_k(X_k) > 
v_i(X_i) + v_j(X_j) + \sum_{k\neq i,j} v_k(X_k)= V(X). 
\]
However, this contradicts the fact that $X$ is an unweighted 
social welfare maximizing allocation. 
From Lemma~\ref{lem:max-value}, it follows that $X$ is weighted envy-freeable. 
\end{proof}

\begin{definition}[VCG mechanism~\citep{clarke,groves:econometrica:1973,vickrey:1961}]
The VCG chooses allocation $X=X^{N,M}$. 
Agent $i$, who is allocated $X_i$, 
pays the amount $$V(X^{N\setminus\{i\}, M}) 
- V(X^{N\setminus\{i\}, M\setminus X_i}).$$ 
\end{definition}
\begin{theorem}
\label{thm:supermodular_VCG}
When valuations are super-modular, 
the VCG mechanism with a large up-front subsidy 
(i.e.,we first distribute $C\cdot w_i$ to agent $i$, and if agent $i$ obtains 
a bundle, it pays the VCG payment from $C\cdot w_i$) is weighted envy-free, 
Pareto efficient, and truthful. 
\end{theorem}
A similar mechanism is presented in \citet{Goko2024} for the unweighted case.

\subsection{Additive valuation}

In the unweighted case
for general additive valuations,
we can always find an envy-freeable allocation with a subsidy at most one for each agent using the ``iterated matching algorithm'' \citep{Brustle2020}.
% proposed by \citet{Brustle2020}.
However, in the weighted case, worst-case subsidy bounds deteriorate 
even in more restricted subclasses, as shown in this subsection. 
In the following, we examine worst-case 
subsidy bounds for general additive valuations, as well as representative subclasses: identical valuations, binary valuations, and identical items.

\subsubsection{General additive valuation}
\begin{proposition}
\label{prop:general-additive}
    There exists a polynomial time algorithm to find a weighted envy-freeable and Pareto efficient %non-wasteful
    allocation with the subsidy bound $m\frac{w_{max}}{w_{min}}$.
\end{proposition}

A mechanism that uses the above algorithm with the second price and 
a large up-front subsidy is truthful; it is one instance of the VCG. 
Note that the subsidy bound for the above algorithm is tight. 
%For subsidy bound, 
Let us assume there are two agents: 
agent 1 with $w_{min}$ and agent 2 with $w_{max}$, respectively, and $m$ items. 
Agent 1 values each item as $1$, while agent 2 values each item as $1-\epsilon$. 
The algorithm allocates all items to agent 1. The subsidy for agent 2 
must be $m(1-\epsilon)w_{max}/w_{min}$.

\subsubsection{Identical additive valuation}
This subsection deals with the case where all agents have identical valuations.
%In this case, 
For each agent $i\in N$ valuation function $v_i$ is denoted by $v$ since they are all the same.
We then show that in this case, the %worst-case 
upper and lower bounds of the subsidy for each agent coincide at one.

It is important to note that with identical valuations, any allocation is weighted envy-freeable since it satisfies weighted reassignment-stability.
Furthermore, all allocations are non-wasteful. %for identical additive valuations.
WLOG, we assume $v(g) > 0$ for each item $g$.

We first show the lower bound for the subsidy. %and the proof is in Appendix.
\begin{theorem}\label{thm:subsidy-lb-identical-additive}
For identical additive valuation, there exists an instance where, for any weighted envy-freeable 
% and non-wasteful 
allocation, at least one agent requires a subsidy of
one.
\end{theorem}

We then show the upper bound.
In fact, the following simple algorithm outputs an allocation s.t. the subsidy is at most one.

\begin{algorithm}
\caption{Polynomial time algorithm for one unit of subsidies}\label{alg:one-subsidy-identical-additive}
\begin{algorithmic}[1]
\REQUIRE Allocation $X$ is empty at the beginning
%\ENSURE $y = x^n$
\FOR{$g:$ 1 to $m$}
\STATE $u=\arg\min_{i\in N}\frac{v(X_i \cup \{g\})}{w_i}$
\STATE Add $g$ to $X_u$
\ENDFOR
\end{algorithmic}
\end{algorithm}

For the output of Algorithm \ref{alg:one-subsidy-identical-additive}, we have the following lemma.
\begin{lemma}\label{lem:length-bound-identical-additive}
    For any path $P$ starting from agent $i$, $\ell(P)\le \frac{1}{w_i}$ holds.
\end{lemma}

\begin{theorem}\label{thm:poly-on-subsidy-identical-additive}
For identical additive valuation, there exists a polynomial time algorithm to find an envy-freeable and non-wasteful allocation such that the subsidy for each agent is at most one.
\end{theorem}
\begin{proof}
%The polynomial time algorithm is Algorithm \ref{alg:one-subsidy-identical-additive}. 
It is clear that the Algorithm~\ref{alg:one-subsidy-identical-additive} runs in polynomial time. 
As noted at the beginning of this subsection, it is also non-wasteful.

Next, we analyze the output of Algorithm \ref{alg:one-subsidy-identical-additive}.
Let $i$ be any agent. Suppose that $P$ is a longest weighted path from $i$ in the weighted envy-graph. 
Then we have $\ell(P)\le \frac{1}{w_i}$ by Lemma~\ref{lem:length-bound-identical-additive}.
%lemmas \ref{lem:balanced-identical-additive} and \ref{lem:pathbound-identical-additive}. 
So, the subsidy for $i$ is bounded by one from Theorem~\ref{thm:minsubsidy}.
\end{proof}

\subsubsection{Binary additive valuation}
%%% Assume $v_i(g) \in \{0, 1\}$ for any $i\in N$ and $g \in M$. 
In this subsection, 
we assume $v_i(g) \in \{0,1\}$ for all $i \in N$ and $g \in M$.
WLOG, we assume 
for each $g \in M$, there exists at least one agent $i$ s.t. 
$v_i(g)=1$ holds (otherwise, $g$ is useless for every agent and we may 
remove $g$ from $M$ without affecting weighted envy-freeness). 
Then, non-wastefulness is equivalent to the fact that 
for each agent $i$, $v_i(X_i)= |X_i|$ holds. 

\begin{observation}
For binary additive valuations, any non-wasteful allocation is Pareto efficient and maximizes unweighted social welfare.
Since binary additive valuations are a subclass of super-modular valuations, it follows that any non-wasteful allocation is weighted envy-freeable.
\end{observation}

\begin{theorem}\label{thm:subsidy-lb-binaryadditive}
For binary additive valuation, there exists an instance where, for any weighted envy-freeable %and non-wasteful 
allocation, at least one agent requires a subsidy of
$\frac{w_{max}}{w_{min}}$. 
\end{theorem}
\begin{proof}
Assume there are three agents $1, 2, 3$ with weights $w_{max}$, 
$w_{min}$, $w_{min}$, respectively. 
There is only one item $g$, where $v_1(g)=0$ and $v_2(g)=v_3(g)=1$. 
In a weighted envy-freeable allocation, the item must be allocated to 
either $2$ or $3$. WLOG, assume the item is allocated to $2$. 
Then, we need to pay $1$ for the agent $3$ to eliminate her envy toward $2$. 
Then, we need to pay $\frac{w_{max}}{w_{min}}$ for agent $1$ 
to eliminate her envy toward $3$. 
\end{proof}

In the following, we will show a polynomial time algorithm to reach an allocation that requires at most $w_{max}/w_{min}$ subsidies for each agent. 
Let us first introduce a new notion called \emph{positive path}.
\begin{definition}[Positive path]
    Given  an allocation $X$, we say that there is a positive path from agent $i$ to agent $j$, if there is a path $\{i_0,i_1,\dots,i_k\}$ such that: (1) $i_0=i$ and $i_k=j$; (2) $v_{i_t}(X_{i_{t+1}})>0$ for $t\in \{0,1, \ldots ,k-1\}$.

For simplicity of notation, we assume there is a positive path from $i$ to $i$.  
\end{definition}

\begin{algorithm}
\caption{Polynomial time algorithm for $w_{max}/w_{min}$ subsidies}\label{alg:small-subsidy}
\begin{algorithmic}[1]
\REQUIRE Allocation $X$ is empty at the beginning
%\ENSURE $y = x^n$
\FOR{$g:$ 1 to $m$}
\STATE $V=\{i\mid v_i(g)=1\}$
\STATE $R=\{i\mid \text{there is a positive path from }i\text{ to }j\in V \}$\label{line: R}
\STATE $u=\arg\min_{i\in R}\frac{v_i(X_i)+1}{w_i}$
\STATE Find a positive path from $u$ to $j\in V$
\STATE Add $g$ to $X_j$
\STATE Along the path from $u$ to $j$, for each edge $(i_t,i_{t+1})$ in the path, find an item $g'\in X_{i_{t+1}}$ such that $v_{i_t}(g')=1$
\STATE Move $g'$ from bundle $X_{i_{t+1}}$ to bundle $X_{i_t}$
\ENDFOR

\end{algorithmic}

\end{algorithm}

\begin{lemma}\label{lem:path-weight-1/w}
    On binary additive instance, Algorithm \ref{alg:small-subsidy} will return an allocation such that if there is a positive path from $i$ to $j$, then we have $\frac{v_i(X_i)+1}{w_i}\ge\frac{v_j(X_j)}{w_j}$.
\end{lemma}

\begin{theorem}\label{thm:poly-small-subsidy}
    There exists a polynomial time algorithm to find an envy-freeable and non-wasteful allocation such that the subsidy for each agent is at most $\frac{w_{max}}{w_{min}}$.
\end{theorem}
\begin{proof}
    %The polynomial time algorithm is Algorithm \ref{alg:small-subsidy}. 
    We give a short proof that Algorithm~\ref{alg:small-subsidy} is polynomial time. First, the number of for loops is bounded by $m$. Inside the loop, we can use breadth-first-search (BFS) to identify the set $R$ efficiently. All remaining computations can be performed efficiently.

    Next, we analyze the output of Algorithm \ref{alg:small-subsidy}.
 As for any $h$ and $k$, we have $v_h(X_k)\le v_k(X_k)$ since the allocation is non-wasteful. Thus, from Lemma~\ref{lem:max-value},     
        for any path $P$ from $i$ to $j$ in the weighted envy-graph, we have $\ell(P) \le \frac{v_j(X_j)}{w_j}-\frac{v_i(X_i)}{w_i}$. 
 
    Let $i$ be any agent. Suppose that $P$ is the longest weighted path from $i$ in the weighted envy-graph. 
    If path $P$ is a positive path, then we have $\ell(P)\le \frac{1}{w_i}$ by Lemma \ref{lem:path-weight-1/w}. If path $P$ is not a positive path, then we can divide it into three consecutive parts $P_1$, $P_2$ and $P_3$ such that  $P_3$ is a positive path and $P_2$ contains only one edge $(k,h)$ such that $v_k(X_h)=0$. By %Claim \ref{claim:pathbound}, 
    Lemma~\ref{lem:max-value},
    we have $\ell(P_1)\le \frac{v_k(X_k)}{w_k}-\frac{v_i(X_i)}{w_i}$. By Lemma \ref{lem:path-weight-1/w}, we have $\ell(P_3)\le 1/w_h$. For path $P_2$, we have $\ell(P_2)=-\frac{v_k(X_k)}{w_k}$. Sum them together we have $\ell(P)\le \frac{1}{w_h}-\frac{v_i(X_i)}{w_i}$.

    Thus, any path length is bounded by $1/w_{min}$. So, the subsidy is bounded by $w_{max}/w_{min}$.
\end{proof}

\begin{corollary}
Consider the allocation returned by Algorithm \ref{alg:small-subsidy}. For any agent $i$, if  $v_i(X_i)\ge w_i/w_{min}-1$, then the subsidy of $i$ is bounded by 1.
\end{corollary}
\begin{proof}
    There are two bounds of the longest path of weighted envy-graph in the proof of Theorem \ref{thm:poly-small-subsidy}: 1) $\frac{1}{w_i}$; 2) $\frac{1}{w_{min}}-\frac{v_i(X_i)}{w_i}$. If we have $v_i(X_i)\ge w_i/w_{min}-1$, then both bounds converge to $\frac{1}{w_i}$. The subsidy is bounded by $w_i\cdot \frac{1}{w_i}=1$.
\end{proof}

Somewhat surprisingly, if the binary condition is violated even slightly, Theorem~\ref{thm:poly-small-subsidy} ceases to hold even in settings very close to the unweighted case.
See Theorem~\ref{thm:subsidy-lb-additive-identical-items} and its proof for details.

\textbf{Matroidal valuation.} Another valuation that is slightly more general than binary additive 
would be a matroidal valuation~\citep{BarmanVermaAAMAS2021,benabbou2020}.
A matroidal valuation, which is based on a rank function of a matroid \citep{oxley2011matroid}, is defined as follows. 
 For each agent, we have ${\mathcal F}$, where $F \subseteq M$ holds for each $F \in {\mathcal F}$. 
We assume ${\mathcal F}$ is a matroid, i.e., $\emptyset \in {\mathcal F}$, $F' \subseteq F \in {\mathcal F}$ implies $F' \in {\mathcal F}$, and for any $F, F' \in {\mathcal F}$ where 
$|F|>|F'|$, there exists an item $g \in F\setminus F'$ s.t. $F'\cup\{g\} \in {\mathcal F}$. 
Then, $v_i(A) = \max_{F \in {\mathcal F}} |A\cap F|$. 
A binary additive valuation is a special case of a matroidal valuation;
we can assume agent $i$ has ${\mathcal F}= \{F \mid F\subseteq B\}$,
where $B=\{g \mid g \in M, v_i(g)=1\}$.
Note that a matroidal valuation is no longer additive; it is \emph{sub-modular}.

The following theorem shows that 
we cannot bound the subsidy % for each agent 
by $w_{max}/w_{min}$, assuming $m\geq 3$ and $w_{max}/w_{min}\geq \frac{m}{m-2}$.
\begin{theorem}\label{thm:subsidy-lb-matroidal}
For matroidal valuations, there exists an instance such that, for any weighted envy-freeable and non-wasteful allocation, 
the minimum subsidy required is $\frac{m}{2}\cdot 
(\frac{w_{max}}{w_{min}}-1)$.  

\end{theorem}
\begin{proof}
Assume there are two agents 1 and 2, with weights $w_{max}$, $w_{min}$, respectively. 
There are $m=2k$ items. The valuation of each agent $i$ for bundle $X_i$ 
is given as: $\min(k, |X_i|)$, i,e., each agent needs at most $k$ items
(we can assume ${\mathcal F} = \{F \mid F\subseteq M, |F|\leq k\}$). 
Then, the only non-wasteful allocation allocates $k$ items for each agent.
The edge from agent 1 to 2 % in the envy-graph 
has length
$k/w_{min} - k/w_{max}$. 
Thus, we need to pay $k(w_{max}/w_{min} -1)$ to agent 1. 
\end{proof}

\subsubsection{Identical items}
This subsection deals with the case where the items are identical with additive valuations.
In this scenario, each agent's valuation depends only on the number of items allocated to her.
We show %in this case 
upper and lower bounds of the subsidy for each agent that are almost identical: upper bound $(n-1)\frac{w_{max}}{w_{min}}+1$ and lower bound $(n-1)\frac{w_{max}}{w_{min}}$, where $n$ is the number of agents.

Because of the additivity and identity assumptions, the valuation of agent $i$ is a constant multiple of the number of items allocated to her.
With a slight abuse of notation, we also denote this constant by $v_i$.
Thus, if $m_i$ items are allocated to agent $i$ then her valuation for this allocation is $v_i \cdot m_i$.
WLOG, we assume that $0 \leq v_1 \le v_2 \le \ldots \le v_n (\le 1)$.
For simplicity, we denote an allocation by a tuple of numbers of items allocated to the agents: $(m_1,m_2,\ldots, m_n)$, where $m_i$ is the number of items allocated to agent $i$.
Note that $m = \sum_{i \in N}m_i$ holds.

The following useful lemma states that weighted envy-freeablity can be characterised by the weighted reassignment-stability condition for swapping only a pair of two agents.
% in the case of identical items.
\begin{lemma}
\label{lem:WEFability-additive-identical-items}
For additive valuation with identical items, 

an allocation $(m_1,m_2\dots, m_n)$ is weighted envy-freeable if and only if 
for each $1 \le i,j \le n$ with $v_i < v_j $ we have 
$\frac{m_i}{w_i} \le \frac{m_j}{w_j}$ (or, equivalently, $\frac{v_i \cdot m_i}{w_i} + \frac{v_j \cdot m_j}{w_j} \ge \frac{v_i \cdot m_j}{w_j} + \frac{v_j \cdot m_i}{w_i}$).
\end{lemma}

\begin{theorem}\label{thm:subsidy-lb-additive-identical-items}
For $n(\ge 2)$ agents with additive valuations and identical items, 
the subsidy for an agent can be $(n-1)\cdot \frac{w_{max}}{w_{min}}-\varepsilon$ for any $\varepsilon > 0$, 
assuming there exist $n(n-1)/2$ items.
\end{theorem}

The above lower bound also holds for additive valuations in general, assuming $m\geq n(n-1)/2$. 

We then show the upper bound on the subsidy.

\begin{algorithm}
\caption{Polynomial time algorithm for $(n-1)\frac{w_{max}}{w_{min}}+1$ subsidies}\label{alg:n-subsidy-additive-identical-items}
\begin{algorithmic}[1]
\REQUIRE Allocation $X$ is empty at the beginning
\STATE Sort the agents in an order such that if $v_i<v_j$, then $i<j$. Fix such order
%\ENSURE $y = x^n$
%\STATE Let $R = \{ (i,j) \in N^2 \mid v_i < v_j \}$
\FOR{$g:$ 1 to $m$}
\STATE Let $N'=\{i\in N\mid \frac{m_i+1}{w_i}\le \frac{m_{i+1}}{w_{i+1}}\}$ (We always have $n\in N'$)
%Let $N' \subseteq N$ be the set of agents such that allocation of $g$ to her does not violate weighted envy-freeability, i.e., $N' = \{ i \in N \mid \text{$(X_1,\dots, X_i\cup \{g\},\dots, X_n)$ is weighted envy-freeable} \}$
\STATE %$u=\arg\min_{i\in N'}\frac{v(X_i \cup \{g\})}{w_i}$ \xin{I could prove if $u=\min\{i\in N'\}$. I guess they are equivalent}
Let $u=\min_{i \in N'} i$
\STATE Add $g$ to $X_u$
\ENDFOR
\end{algorithmic}
\end{algorithm}

\begin{theorem}
\label{thm:subsidy-ub-additive-identical-items}
Algorithm \ref{alg:n-subsidy-additive-identical-items} will output an allocation with the subsidy bound $(n-1)\frac{w_{max}}{w_{min}}+1$ for each agent.
\end{theorem}
\begin{proof}
Note that the allocation output by Algorithm \ref{alg:n-subsidy-additive-identical-items} is weighted envy-freeable by Lemma~\ref{lem:WEFability-additive-identical-items}.
We prove that while running the algorithm, 
the length of the longest path is always bounded by $\sum_{1\le k\le n} 1/w_k$. 

From the proof of Lemma~\ref{lem:WEFability-additive-identical-items}, WLOG, we can assume the longest path 
is $1\rightarrow 2 \rightarrow \ldots \rightarrow n$, since $\ell(i, i+1)$ is non-negative, $\ell(i, j)$ where $i>j$ is non-positive, and there exists no positive cycle. 
Also, $\ell(i, j)$ where $j\geq i+2$ is smaller than or equal to $\sum_{i\leq k < j} \ell(k, k+1)$. 

Now, we prove $\sum_{1\le k < n} \ell(k,k+1) \le \sum_{i\le k\le n} 1/w_k$ by induction. 
When there is no allocated item, this must be true. 
Now, suppose that the algorithm will allocate an item to agent $u$. 

When $u=n$, by the fact that the algorithm chose $n$, before the allocation, 
we should have  $\frac{m_i+1}{w_i}>\frac{m_{i+1}}{w_{i+1}}$ for each $i<n$. Otherwise, the algorithm should choose agent $i$ to allocate the item. 
Thus, $\ell(i,i+1)$ is bounded by $1/w_i$ for each $i<n$ before the allocation. 
After allocating the item to agent $n$, the path length increases by $1/w_n$. 
Thus, the new path length is at most $\sum_{1\le k < n} 1/w_k+ 1/w_n$. 

When $u<n$, $\ell(u-1, u)$ increases by $v_{u-1}/w_u$, while $\ell(u, u+1)$ decreases by $v_{u}/w_u$. 
The lengths of other edges are the same. 
Thus, the total path length weakly decreases. From the induction assumption, the path length is bounded by $\sum_{1\le k\le n}1/w_k$. 
Now we can conclude that the length of a longest path is bounded by $\sum_{1\le k\le n}1/w_k$. 
Thus, the subsidy for agent $i$ is at most $$w_i\left(\sum_{1\le k\le n}1/w_k\right)\le (n-1)\frac{w_{max}}{w_{min}}+1.$$
\end{proof}

\section{Approximate envy-free allocations that are weighted envy-freeable}
\label{sec:app_EF}
% \saha{In this section, allocation is denoted by $A_i$, we need to change to $X_i$.}

In this section, we try to achieve weighted envy-freeability along with weighted EF1 type properties.

\begin{definition}
An allocation $X$ is said to be \emph{(strongly) weighted envy-free up to one item} (WEF1) if for any pair of agents $i,j$ with $X_j\neq\emptyset$, there exists an item $g \in X_j$ such that \[\frac{v_i(X_i)}{w_i} \ge  \frac{v_i(X_j \setminus \{g\})}{w_j}.\]

\end{definition}

\begin{definition}
An allocation $X$ is said to be \emph{weakly weighted envy-free up to one item} (WWEF1) if for any pair of agents $i,j$ with $X_j\neq\emptyset$, there exists an item $g \in X_j$ such that 
\[\text{either } \frac{v_i(X_i)}{w_i} \ge  \frac{v_i(X_j \setminus \{g\})}{w_j} \text{ or } \frac{v_i(X_i\cup\{g\})}{w_i} \ge  \frac{v_i(X_j)}{w_j}.\]
\end{definition}

By definition, if an allocation is WEF1, then it is WWEF1.

We focus on additive valuations in what follows in this section and examine the question of whether WEF1 and weighted envy-freeability are compatible. 
Our first observation is that even for 2 agents, the Weighted Picking Sequence Protocol~\citep{CISZ21}, which outputs a WEF1 allocation for any number of agents with additive valuations, does not satisfy weighted envy-freeability. 

\begin{example}

Suppose there is one item $g$ and $v_1(g)=1$, $v_2(g)=2$, $w_1=4/5$, and $w_2=1/5$. 
Agent 1 gets the first turn and gets $g$. However,  this allocation is not envy-freeable since it does not satisfy weighted reassignment-stability. 
\end{example}

In the unweighted case, 
for additive valuations 
we can always find an envy-freeable EF1 allocation by the ``iterated matching algorithm'' proposed by %Brustle et al.
\citet{Brustle2020}.

In the weighted case, there exists an instance such that there exists no weighted envy-freeable and WWEF1 allocation.

\begin{theorem}\label{thm:additive-incompatibility-EFable-WWEF1}
There may not exist weighted envy-freeable allocation that is WWEF1 even for 2 agents, 2 identical items with agents having additive valuations. 
\end{theorem}
% \begin{proof}
%     The proof is in Appendix.
% \end{proof}

For this reason, we now consider a weaker notion. 
\begin{definition}[Weighted envy-free up to one item transfer (WEF1-T)~\citep{AGM23a,HSV23a}]
An allocation $X$ is said \emph{weighted envy-free up to one item transfer} (WEF1-T) if for any pair of agents $i,j$ with $X_j\neq\emptyset$,
either $\frac{v_i(X_i)}{w_i} \ge  \frac{v_i(X_j)}{w_j}$ or there exists some $g\in X_j$ such that
$\frac{v_i(X_i\cup\{g\})}{w_i} \ge  \frac{v_i(X_j\setminus\{g\})}{w_j}$.
\end{definition}

Next, we present a particular version of Weighted Adjusted Winner~\citep{CISZ21}, which needs to make decisions differently to get the desired axiomatic properties we are after. 
Here, Weighted Adjusted Winner is a protocol that finds a WEF1 and Pareto efficient allocation for two agents with additive valuations.
We call the rule Biased Weighted Adjusted Winner Procedure that is biased towards the agent who expresses a higher value for a `contested' item.

\paragraph{\textbf{Biased Weighted Adjusted Winner Procedure:}}
We normalize the valuations so that the sum of values over all items is the same for both agents. 
We sort items as $\frac{v_1(g_1)}{v_2(g_1)} \ge \frac{v_1(g_2)}{v_2(g_2)} \ge \dots \ge \frac{v_1(g_m)}{v_2(g_m)}$.

We want to place a boundary so all the items left of the boundary are given to agent 1, and all the items right of the boundary are given to agent 2.
Concretely, let $d \in \{1,2,...,m\}$ be a number satisfying $\frac{1}{w_1}\sum_{r=1}^{d-1}v_1(g_r) < \frac{1}{w_2}\sum_{r=d}^{m}v_1(g_r)$ and $\frac{1}{w_1}\sum_{r=1}^{d}v_1(g_r) \ge \frac{1}{w_2}\sum_{r=d+1}^{m}v_1(g_r)$.
We then see if it is possible to split the items so that we reach WEF.

If yes, we are done. If not, we get a fractional allocation $X$ by cutting the contested item $g_d$ fractionally to ensure WEF. We give it to the agent who has a higher value for it to ensure that the obtained allocation is weighted envy-freeable.

\begin{theorem}
The outcome of the Biased Weighted Adjusted Winner Procedure satisfies WEF1-T and weighted envy-freeability.
\end{theorem}
\begin{proof}
We first prove the algorithm is guaranteed to achieve a fractional WEF allocation that splits at most one item. For the case of 2 agents, WEF is equivalent to the weighted proportionality (WPROP) that requires that each agent $i$ gets a value that is at least $v_i(M) w_i$.

We know that if items are ordered according to the algorithm but agent 2's valuation function is $v_1$, then
there is a unique way to split at most 1 item to get a WPROP (equivalently WEF) outcome for agents 1 and 2. Now consider changing the valuation of agent 2 from $v_1$ to $v_2$. The leftmost boundary to achieve WPROP for agent 1 does not move but the rightmost boundary for agent 2 to achieve WPROP can only move right as more weight is transferred right. Hence, there exists a WEF allocation that can be achieved by splitting up at most one item.

If there is no contested item, the allocation is WEF and hence WEF1-T. Consider the case that there is a contested item. If the contested item $g_d$ is transferred to the envious agent, the envious agent is not envious any more.  Hence, the outcome is WEF1-T.

Next, we prove that the outcome is weighted envy-freeable. If there is no contested item, then the envy-graph has no positive weight, so the allocation is weighted envy-freeable. The second case is that there is a contested item $g_d$.
 $L$ is allocated to agent 1
and $R$ is allocated to agent 2 while excluding $g_d$. Furthermore, let us assume 
$g_d$ can be divided into two items $g_{d_l}$ and $g_{d_r}$; 
in the fractional allocation, $g_{d_l}$ is allocated to agent 1 and 
$g_{d_r}$ is allocated to agent 2 in a WEF manner.  
Let us assume $g_d$ is allocated to agent 1. 
Then, $v_1(g_{d_l}) \geq v_2(g_{d_l})$ and 
$v_1(g_{d_r}) \geq v_2(g_{d_r})$ hold. 
Since the fractional allocation is weighted envy-free, 
$(v_1(L) + v_1(g_{d_l}))/w_1 \ge (v_1(R) + v_1(g_{d_r}))/w_2$ and 
$(v_2(R) + v_2(g_{d_r}))/w_2 \ge (v_2(L) + v_2(g_{d_l}))/w_1$ holds. 
We show that when allocating $g_d$ to agent 1, 
reassignment stability is satisfied, i.e., 
$(v_1(L) + v_1(g_{d}))/w_1 + v_2(R)/w_2 \ge 
 v_1(R)/w_2 + (v_2(L) + v_2(g_d))/w_1$ holds. 
We obtain: 
$(v_1(L) + v_1(g_{d}))/w_1 + v_2(R)/w_2 
 - v_1(R)/w_2 - (v_2(L) + v_2(g_d))/w_1
 \geq (1/w_1 + 1/w_2) (v_1(g_{d_r}) - v_2(g_{d_r})) \ge 0$. 
The case that $g_d$ is allocated to agent 2 can be proved in a similar way. 
\end{proof}

It remains open whether weighted envy-freeability and WEF1-T are compatible for any number of agents. 

\section{Monetary envy-freeness (MEF)}
\label{sec:MEF}

{To achieve weighted envy-free, sometimes it costs too much subsidy. However, in practice, the subsidy could be very limited. A natural question here is: how do we achieve partial fairness when we only have limited subsidies? A possible solution is to first give the subsidy to those whom nobody envies and do not create more envy due to the subsidy. Motivated by this consideration, we define the following concept.}
\begin{definition}
	We say that an outcome $(X,p)$ is \emph{monetarily envy-free (MEF)} if $(X,p)$ is weighted envy-free or it is the case that if for any two agents $i,j\in N$ such that $i$ weighted envies $j$, then $p_j=0$.
	\end{definition}

	Note that the definition bypasses the issue of whether the indivisible item allocation is fair or not. It ensures that the money is allocated in a way to respect a fairness condition. Next, we present an extended definition of weighted envy-graph that also considers payments.

	For any given allocation $X$, the corresponding \emph{weighted envy-graph respecting the subsidy}
	is a complete directed graph with vertex set $N$. For any pair of agents $i,j\in N$ the length of edge $(i,j)$ 
	is the envy agent $i$ has for agent $j$ under $(X,p)$:  $\ell(i,j) \ =\  \frac{1}{w_j}(v_i(X_j)+p_j)-\frac{1}{w_i}(v_i(X_i)+p_i)$. For any path or cycle $C$ in the graph, the length of the $C$ is the sum of the lengths of edges along $C$.

Next, we show that we can achieve MEF using any weighted envy-freeable allocation. 

\begin{theorem}
If an allocation is weighted envy-freeable, then we can allocate any amount of money in a way so that the outcome is MEF. 
\end{theorem}
\begin{proof}

	Let $\ell_i$ be the maximum length of any path in the weighted envy-graph that starts from $i$. 
	If the total money $d$ is at exactly $\sum_{i\in N}\ell_i \cdot w_i$ let each agent $i$'s subsidy be $p_i=\ell_i \cdot w_i$. If $d>\sum_{i\in N}\ell_i\cdot w_i$, we then first allocate the money $\sum_{i\in N}\ell_i\cdot w_i$ so that $i$'s payment be $p_i=\ell_i\cdot w_i$ in which case the outcome is WEF according to the characterization. 
 The surplus amount is paid in proportion to the weights of the agents, which is WEF as well.

	The last case is if $d<\sum_{i\in N}\ell_i \cdot w_i$. In that case, we identify the set of agents $N^*$ who have the highest $\ell_i$. Note that there is no agent $j$ outside $N^*$ who has zero or positive edge to any agent in $N^*$ because if this is the case, $j$ would be in $N^*$. Therefore, any agent in $i\in N^*$ can be given a tiny amount of money without leading some agent outside $N^*$ to become envious. No agent in $i\in N^*$ has a strictly positive edge to $k\in N^*$ or else $k$ would not be a part of $N^*$. When we allocate the money in proportion to the weights, all the $\ell_i$ for $i\in N^*$ decrease at the same rate. As we do this, the set $N^*$ may increase. Eventually, all the money is allocated. 
	\end{proof}

    For additive valuations and unweighted agents and money, there is a simple algorithm to compute an MEF outcome. We use the algorithm of %Brustle et al. 
    \citet{Brustle2020}
    to compute an allocation that is both EF1 and envy-freeable. After that, we use the algorithm above to allocate the money in an MEF way. The outcome is MEF. 

\section{Conclusions}
\label{sec:conclusions}

Envy-free allocation with monetary subsidies is a fundamental problem. In this paper, we examine the topic of when agents can have asymmetric entitlements. 
We showed that weighted entitlements pose new challenges, and various results do not extend from unweighted to weighted entitlements. We present both contrasting results and also certain generalizations. 
It remains open whether weighted envy-freeability and WEF1-T are compatible for any number of agents.
Also, as shown in Table~\ref{tbl:results}, 
for additive and matroidal valuations, there exist large gaps between the lower and upper bounds of the subsidy. Possible future work includes narrowing/eliminating these gaps. 

\paragraph{Acknowledgements.}
This work was supported by the NSF-CSIRO grant on “Fair Sequential Collective Decision-Making" (RG230833). The authors affiliated with Kyushu University are partially supported by JSPS KAKENHI Grant Number JP21H04979 and JST ERATO Grant Number JPMJER2301.

\section*{Appendix}
\label{sec:Appendix}
%\subsection*{Omitted Proofs}
\newtheorem*{thm:supermodular_VCG}{Theorem~\ref{thm:supermodular_VCG}}
\begin{thm:supermodular_VCG}
%\label{thm:supermodular_VCG}
When valuations are super-modular, 
the VCG mechanism with a large up-front subsidy 
(i.e.,we first distribute $C\cdot w_i$ to agent $i$, and if agent $i$ obtains 
a bundle, it pays the VCG payment from $C\cdot w_i$) is weighted envy-free, 
Pareto efficient, and truthful. 
\end{thm:supermodular_VCG}
\begin{proof}
Truthfulness and Pareto efficiency are clear. 
We show that it is weighted envy-free. 
We first show that in the VCG, 
for each agent $i$ who obtains $X_i$ and pays $q_i$, 
$q_i \geq v_j(X_i)$ holds for any $j\neq i$. 
For the sake of contradiction, 
assume $q_i=  
V(X^{N\setminus\{i\}, M}) 
 - V(X^{N\setminus\{i\}, M\setminus X_i})
< v_j(X_i)$ holds. 
Then, 
$V(X^{N\setminus\{i\}, M}) 
< v_j(X_i)  + V(X^{N\setminus\{i\}, M\setminus X_i})$ holds. 
However, if we consider an allocation of $M$ to agents except for $i$, 
we can first allocate $M\setminus X_i$ optimally among $N\setminus \{i\}$, 
then allocate $X_i$ additionally to agent $j$. Then, 
the total valuation of this allocation is at least 
$v_j(X_i)  + V(X^{N\setminus\{i\}, M\setminus X_i})$ due to 
super-modularity. This contradicts the fact that 
$V(X^{N\setminus\{i\}, M})$ is the total valuation when allocating 
$M$ optimally among agents except for $i$. 
Also, $v_i(X_i) \geq q_i$ holds for all $i \in N$, i.e., VCG is individually rational, 

If agent $j$ has an envy toward agent $i$, 
$(v_j(X_j) + C\cdot w_j - q_j)/w_j <  
(v_j(X_i) + C\cdot w_i - q_i)/w_i$ holds. 
Then, $(v_j(X_j) - q_j)/w_j <  
(v_j(X_i) - q_i)/w_i$ must hold. 
However, $v_j(X_j) - q_j \geq 0$, and 
$v_j(X_i) - q_i \leq 0$ hold. This is a contradiction. 
\end{proof}

\newtheorem*{prop:general-additive}{Proposition~\ref{prop:general-additive}}
\begin{prop:general-additive}
    There exists a polynomial time algorithm to find a weighted envy-freeable and Pareto efficient %non-wasteful
    allocation with the subsidy bound $m\frac{w_{max}}{w_{min}}$.
\end{prop:general-additive}

\begin{proof}
Since additive valuation is a special case of super-modular valuation, 
any unweighted social welfare maximizing allocation is weighted envy-freeable. It is also Pareto efficient. 
We can find an unweighted social welfare maximizing allocation using the following polynomial time algorithm: 
for each $g \in M$, allocate it to agent $i$ such that $v_i(g)$ is largest. 
Clearly, $v_i(X_i)\geq v_j(X_i)$ holds for any $i, j \in N$. 
Thus, from Lemma~\ref{lem:max-value}, the subsidy for each agent is bounded 
by $m\frac{w_{max}}{w_{min}}$.
\end{proof}

\newtheorem*{thm:subsidy-lb-identical-additive}{Theorem~\ref{thm:subsidy-lb-identical-additive}}
\begin{thm:subsidy-lb-identical-additive}
For identical additive valuation, there exists an instance where, for any weighted envy-freeable 
% and non-wasteful 
allocation, at least one agent requires a subsidy of
one.
\end{thm:subsidy-lb-identical-additive}

\begin{proof}
Assume there are two agents $1, 2$ with weights $1/2, 1/2$, respectively. 
There is only one item $g$, where $v(g)=1$.
In a non-wasteful allocation, the item must be allocated to 
either agent $1$ or $2$. WLOG, assume the item is allocated to $1$.
Then, we need to pay one for agent $2$ to eliminate her envy toward $1$. 
\end{proof}

\newtheorem*{lem:length-bound-identical-additive}{Lemma~\ref{lem:length-bound-identical-additive}}
\begin{lem:length-bound-identical-additive}
    For any path $P$ starting from agent $i$, $\ell(P)\le \frac{1}{w_i}$ holds.
\end{lem:length-bound-identical-additive}
\begin{proof}
    Assume that the path ends at $h$. Since the valuation is identical, we have $$\ell(P)=\sum_{(k,j)\in P} l(k,j)=\sum_{(k,j)\in P}\frac{v_k(X_j)}{w_j}-\frac{v_k(X_k)}{w_k}=\frac{v(X_h)}{w_h}-\frac{v(X_i)}{w_i}.$$
    By induction, we can prove this is bounded by $\frac{1}{w_i}$. At the beginning of the algorithm, it is obviously true. During the algorithm, if we add item $g$ to $h$, then we should have $\frac{v(X_i\cup g)}{w_i}\ge \frac{v(X_h\cup g)}{w_h}$. So, after adding item $g$ to agent $h$, the length of the path $P$ is bounded by $\frac{v_i(X_h)}{w_h}-\frac{v_i(X_i)}{w_i}\le \frac{v(g)}{w_i}$. If $g$ is not added to $h$, the length would not increase. So, the length is always bounded by $\frac{1}{w_i}$.
\end{proof}

\newtheorem*{lem:path-weight-1/w}{Lemma~\ref{lem:path-weight-1/w}}
\begin{lem:path-weight-1/w}%\label{lem:path-weight-1/w}
    On binary additive instance, Algorithm \ref{alg:small-subsidy} will return an allocation such that if there is a positive path from $i$ to $j$, then we have $\frac{v_i(X_i)+1}{w_i}\ge\frac{v_j(X_j)}{w_j}$.
\end{lem:path-weight-1/w}

\begin{proof}
    We prove this by induction. This is true with empty allocation. Suppose that before we allocate item $g$, the statement is true. 

    First, let us look at the agents outside the set $R$ (defined in line \ref{line: R}). Let $X'$ denote the allocation after processing line 8. 
    Let $T=\bigcup_{i\in R}X'_i$ be the set of all items in the bundle, which is owned by some agent in the set $R$. 
    % Notice that here $X_i$ is the bundle just after allocation of item $g$. 
    If $i\notin R$, then we have $v_i(T)=0$. Otherwise, agent $i$ could have a positive path to some agent in $R$. No matter what changes we made inside $R$, they do not influence any agent $i\notin R$ in terms of a positive path starting from $i$. Please refer 
    Figure~\ref{fig:binary-additive} for a graphic explanation for these sets.

    Now, we only need to consider how this allocation could influence a positive path starting from agents in set $R$. After the allocation, only agent $u$'s utility increases by 1. We can show that no one would break the statement because of this. Before the allocation, the value $\frac{v_u(X_u)+1}{w_u}$ is the minimum among the set $R$. We have $\frac{v_i(X_i)+1}{w_i}\ge \frac{v_u(X_u)+1}{w_u}=\frac{v_u(X_u')}{w_u}$. Here $X_u'$ is the bundle after the allocation.

    \begin{figure}
        \centering
        \includegraphics[width=1\linewidth]{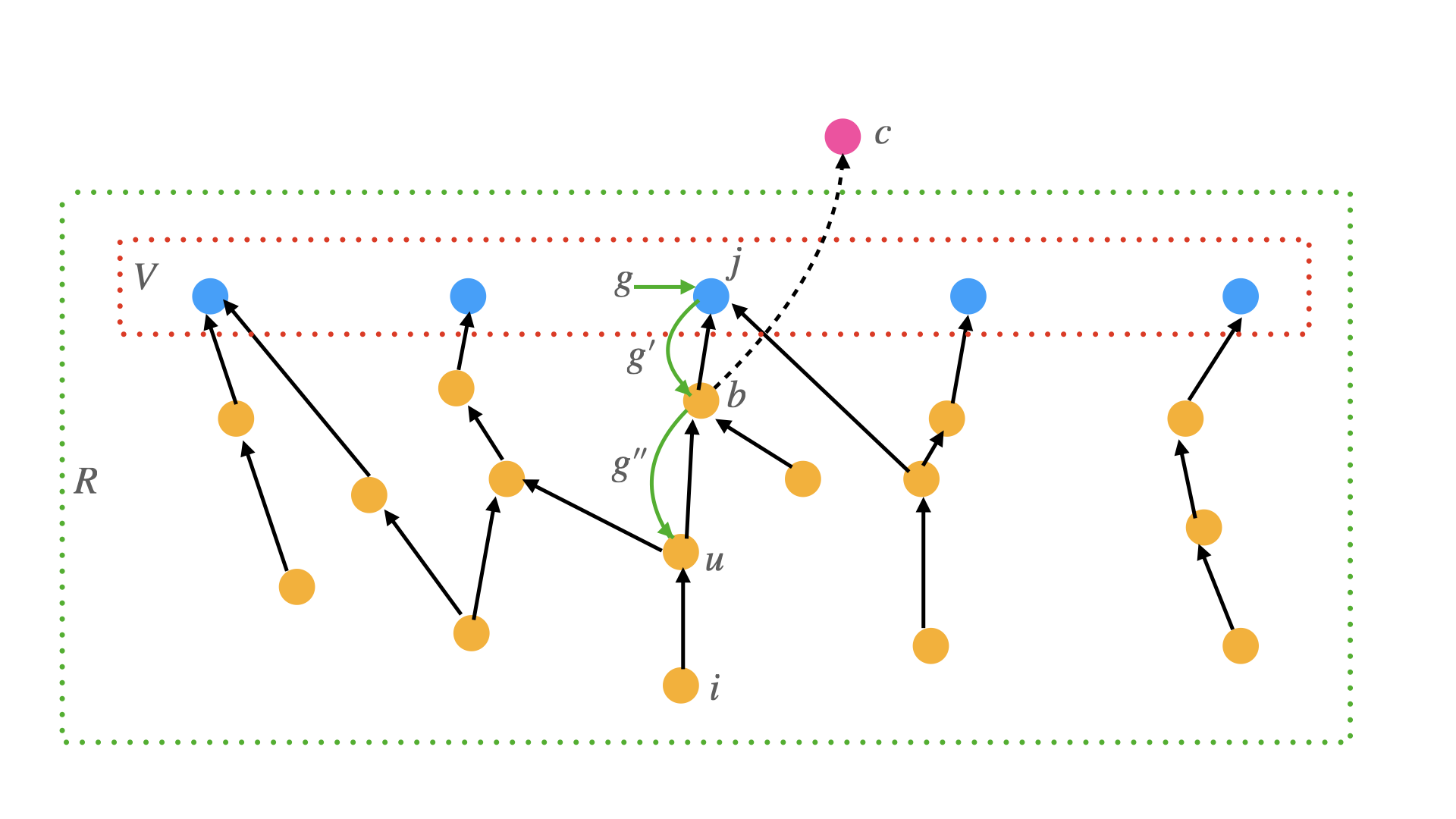}
        \vspace{-0.5cm}
        \caption{A diagram for various variables in the proof of Lemma \ref{lem:path-weight-1/w}.}
        \label{fig:binary-additive}
    \end{figure}

    Even if the utility of all agents except $u$ does not change, the allocation could add a lot of new positive paths and/or increase the length of existing positive paths. Next we show that all affected positive paths satisfying the statement. 
    Notice that only the bundles in the path from $u$ to $j$ change. Let $P$ be the set of agents in the positive path from $u$ to $j$. Then,  only edge $(i_0,i_1)$ where $i_1\in P$ can be changed after the allocation. It could add some new positive paths or increase the length of some existing positive paths. However, those newly affected paths could only end with an agent $c$ such that there is a positive path (before the allocation) from $u$ to $c$ because there is a positive path from $u$ to any agent in set $P$ before the allocation. Note that $c$ can be any agent in $N$. Any affected positive path must be in the form $i$ to some agent $b\in P$ and then end up with some agent $c$. Before the allocation, we have a positive path from $u$ to $b$ and then to $c$. So, by the assumption of our induction,  we should have $\frac{v_u(X_u)+1}{w_u}\ge\frac{v_c(X_c)}{w_c}$. As for any $i\in R$, we have $\frac{v_i(X_i)+1}{w_i}\ge \frac{v_u(X_u)+1}{w_u}\ge\frac{v_c(X_c)}{w_c}$. So for any possible affected positive paths, the statement still holds true.
\end{proof}

\newtheorem*{lem:WEFability-additive-identical-items}{Lemma~\ref{lem:WEFability-additive-identical-items}}
\begin{lem:WEFability-additive-identical-items}
%\label{lem:WEFability-additive-identical-items}
For additive valuation with identical items, 
%assume that $0 < v_1 \le v_2 \le \ldots \le v_n (\le 1)$.
%Then 
an allocation $(m_1,\dots, m_n)$ is weighted envy-freeable if and only if 
for each $1 \le i,j \le n$ with $v_i < v_j $ we have 
$\frac{m_i}{w_i} \le \frac{m_j}{w_j}$ (or, equivalently, $\frac{v_i \cdot m_i}{w_i} + \frac{v_j \cdot m_j}{w_j} \ge \frac{v_i \cdot m_j}{w_j} + \frac{v_j \cdot m_i}{w_i}$).
\end{lem:WEFability-additive-identical-items}

\begin{proof}
We first show the only-if part.
Assume that $(m_1,\ldots, m_n)$ is weighted envy-freeable.
Then, by Theorem~\ref{th:ef}, it is weighted reassignment-stable.
For the permutation that only swaps $i$ and $j$, inequality~\eqref{eq:WRS} in the definition of weighted reassignment-stability implies that 
$\frac{v_i \cdot m_i}{w_i} + \frac{v_j \cdot m_j}{w_j} \ge \frac{v_i \cdot m_j}{w_j} + \frac{v_j \cdot m_i}{w_i}$.
This is equivalent to $\frac{m_i}{w_i} \le \frac{m_j}{w_j}$, since $v_i < v_j$.

We then show the if part.
WLOG, we can assume when $i< j$ and $v_i=v_j$, 
$m_i/w_i \le m_j/w_j$ holds (we can rename agents' identifiers). 
We will show that for any $i < j$ we have $\ell(i,j)\le \ell(i,i+1)+\ell(i+1,i+2)+\dots +\ell(j-1,j)$. %\xin{We can strength this to $\forall \text{path} P_{ij}, \ell(P_{ij})\le \ell(i,i+1)+\ell(i+1,i+2)+\dots +\ell(j-1,j)$.}
Once this is done, we can show that any cycle in the weighted envy-graph has a non-positive length as follows.
Let $C$ be a cycle in the weighted envy-graph.
We partition the set of edges of $C$ into the two sets of ``ascending'' edges and ``descending'' edges: $E(C) = E_+(C) \cup E_-(C)$, where $E_+(C)=\{ (i,j) \in E(C) \mid i<j \}$ and $E_-(C)=\{ (j,i) \in E(C) \mid i<j \}$.
To show $C$ has non-positive length is to show $\sum_{(i,j)\in E_+(C)}\ell(i,j)+\sum_{(j,i)\in E_-(C)}\ell(j,i) \le 0$.
Using the inequality $\ell(i,j)\le \ell(i,i+1)+\ell(i+1,i+2)+\dots +\ell(j-1,j)$, 
$\sum_{(i,j)\in E_+(C)}\ell(i,j)$ is at most $\sum_{(i,j)\in E_+(C)}(\ell(i,i+1)+\ell(i+1,i+2)+\dots +\ell(j-1,j))$.
Let $E'_+(C)$ be a \emph{multiset} of edges of $C$ defined as $E'_+(C)=\cup_{(i,j)\in E_+(C)}\{ (i,i+1),(i+1,i+2),\dots ,(j-1,j) \}$.
Then it suffices to show that $\sum_{(i,j)\in E'_+(C)}\ell(i,j)+\sum_{(j,i)\in E_-(C)}\ell(j,i) \le 0$ holds.
Now, we can disjointly assign edges $(i,i+1),(i+1,i+2),\dots ,(j-1,j) \in E'_+(C)$ to each edge $(j,i) \in E_-(C)$ since $C$ is a cycle.
Then it suffices to show that the sum of the length of ``ascending''  edges assigned to $(j,i) \in E_-(C)$ plus the length of $(j,i)$ (i.e., $\ell(i,i+1)+\ell(i+1,i+2)+\dots +\ell(j-1,j)+\ell(j,i)$) is non-positive, since $\sum_{(i,j)\in E'_+(C)}\ell(i,j)+\sum_{(j,i)\in E_-(C)}\ell(j,i)$ is the sum of these values.
Indeed, we have
\begin{align*}
&\ell(i,i+1)+ \dots +\ell(j-1,j)+\ell(j,i)\\
&=v_{i}\cdot\left(\frac{m_{i+1}}{w_{i+1}}-\frac{m_{i}}{w_{i}}\right)+\dots +v_{j-1}\cdot\left(\frac{m_{j}}{w_{j}}-\frac{m_{j-1}}{w_{j-1}}\right) + v_{j}\cdot\left(\frac{m_{i}}{w_{i}}-\frac{m_{j}}{w_{j}}\right)\\
&\le v_{j}\cdot\left(\frac{m_{i+1}}{w_{i+1}}-\frac{m_{i}}{w_{i}}\right)+\dots +v_{j}\cdot\left(\frac{m_{j}}{w_{j}}-\frac{m_{j-1}}{w_{j-1}}\right) + v_{j}\cdot\left(\frac{m_{i}}{w_{i}}-\frac{m_{j}}{w_{j}}\right)\\
&=v_{j}\cdot\left(\frac{m_{j}}{w_{j}}-\frac{m_{i}}{w_{i}}\right) + v_{j}\cdot\left(\frac{m_{i}}{w_{i}}-\frac{m_{j}}{w_{j}}\right)\\
&= 0,
\end{align*}
where we use the fact that $v_i,\dots, v_{j-1} \le v_{j}$ and $\frac{m_{k}}{w_{k}} \le \frac{m_{k+1}}{w_{k+1}}$ for $i \le k \le j-1$ in the inequality.
Therefore, any cycle in the weighted envy-graph has a non-positive length, and the allocation is weighted envy-freeable by Theorem~\ref{th:ef}.

It remains to show $\ell(i,j)\le \ell(i,i+1)+\ell(i+1,i+2)+\dots +\ell(j-1,j)$ holds.
We show this by induction on $j-i$.
If $j-i = 1$, then $\ell(i,j) = \ell(i,i+1)$ holds.
Assume $j-i > 1$.
By the inductive hypothesis, we have 
$\ell(i,j-1)\le \ell(i,i+1)+\ell(i+1,i+2)+\dots +\ell(j-2,j-1)$.
Hence, it suffices to show that $\ell(i,j)\le \ell(i,j-1)+\ell(j-1,j)$.
Indeed, 
\begin{align*}
\ell(i,j) &= v_{i}\cdot\left(\frac{m_{j}}{w_{j}}-\frac{m_{i}}{w_{i}}\right)\\
&= v_{i}\cdot\left(\frac{m_{j}}{w_{j}}-\frac{m_{j-1}}{w_{j-1}}\right) + v_{i}\cdot\left(\frac{m_{j-1}}{w_{j-1}}-\frac{m_{i}}{w_{i}}\right)\\
&\le v_{j-1}\cdot\left(\frac{m_{j}}{w_{j}}-\frac{m_{j-1}}{w_{j-1}}\right) + v_{i}\cdot\left(\frac{m_{j-1}}{w_{j-1}}-\frac{m_{i}}{w_{i}}\right)\\
&= \ell(j-1,j)+\ell(i,j-1),
\end{align*}
where we use the fact that $v_i \le v_{j-1}$ and $\frac{m_{j-1}}{w_{j-1}} \le \frac{m_{j}}{w_{j}}$ in the inequality.
\end{proof}

\newtheorem*{thm:subsidy-lb-additive-identical-items}{Theorem~\ref{thm:subsidy-lb-additive-identical-items}}
\begin{thm:subsidy-lb-additive-identical-items}%\label{thm:subsidy-lb-additive-identical-items}
For $n(\ge 2)$ agents with additive valuations and identical items, 
the subsidy for an agent can be $(n-1)\cdot \frac{w_{max}}{w_{min}}-\varepsilon$ for any $\varepsilon > 0$.
\end{thm:subsidy-lb-additive-identical-items}

\begin{proof}
Consider the case with $n$ agents with valuation $v_i = 1 - (n-i)\delta$ for $i \in N$ and weights satisfying $w_1=w_{max}$, $w_2=w_{min}$, and $w_i = (1 + \delta)^{i-2}w_{min}$ for $i \ge 3$, where $\delta$ is a sufficiently small positive real number.
There are $m = n(n-1)/2$ items.
%\my{Do you mean $\binom{n+1}{2} -1 = n(n+1)/2 -1$?}
%\kk{Yes, I fixed it.}
For allocation $(m_1,m_2,\dots, m_n)$ to be weighted envy-freeable, it must satisfy $m_{i+1} > m_i$ for $i \ge 2$ by the only-if part of Lemma~\ref{lem:WEFability-additive-identical-items} since $v_i < v_{i+1}$ and $w_{i+1}/w_i = 1 + \delta$.
Also, $m_1 > 0$ holds only if $m_2 > 0$ holds, since $v_1 < v_2$ implies $m_2 \ge (w_{2}/w_{1}) \cdot m_1$ again by Lemma~\ref{lem:WEFability-additive-identical-items}.
Then $m_1$ must be zero, since otherwise $m_i \ge i-1$ must hold for $i \ge 2$ and $\sum_{i \in N}m_i \ge n(n-1)/2 + 1 > m$, a contradiction.
Also, $m_n \ge n-1$ holds since otherwise $m_i \le i-2$ holds for each $i \ge 2$ and $m_1 = 0$ holds, implying that $\sum_{i \in N}m_i \le (n-1)(n-2)/2 < m$, a contradiction.
Now, we claim that the subsidy for agent $1$ is at least $(n-1)\cdot \frac{w_{max}}{w_{min}}-\varepsilon$.
Let us consider the length of the edge from agent 1 to $n$ in the weighted envy-graph.
It is $\frac{1}{w_n}v_n \cdot m_n-\frac{1}{w_1}v_1 \cdot m_1$, which is at least $\frac{1}{w_n}v_n \cdot (n-1)$ since $m_n \ge n-1$ and $m_1=0$ hold.
Therefore, the length of the longest path from $1$ in the weighted envy-graph is at least $\frac{1}{w_n}v_n \cdot (n-1)$ and 
we need to pay $\frac{w_1}{w_n}v_n \cdot (n-1)$ for agent $1$ by Theorem~\ref{thm:minsubsidy}.
By setting $\delta > 0$ sufficiently small, we obtain that 
$\frac{w_1}{w_n}v_n \cdot (n-1)$ is at least $(n-1)\cdot \frac{w_{max}}{w_{min}}- \varepsilon$.
\end{proof}

\newtheorem*{thm:additive-incompatibility-EFable-WWEF1}{Theorem~\ref{thm:additive-incompatibility-EFable-WWEF1}}
\begin{thm:additive-incompatibility-EFable-WWEF1}%\label{thm:additive-incompatibility-EFable-WWEF1}
There may not exist weighted envy-freeable allocation that is WWEF1 even for 2 agents, 2 identical items with agents having additive valuations. 
\end{thm:additive-incompatibility-EFable-WWEF1}

\begin{proof}
Consider the case with two agents $1, 2$, with weights $3/5, 2/5$, respectively.
There are two identical items.
Agent 1 values one item as 120, while agent 2 values one item as 60. 
First, we see that any allocation that allocates one item for each agent does not satisfy weighted reassignment-stability: 
\[\frac{120}{w_1} + \frac{60}{w_2} = 200 + 150 = 350 < 
\frac{120}{w_2} + \frac{60}{w_1} = 300 + 100 = 400.\]
Hence, such an allocation is not envy-freeable.
Then let us consider the other allocations $X = (M, \emptyset)$ and $X' = (\emptyset, M)$.
We show that these allocations are not WWEF1.
It is easy to see that those allocations are not WEF1 since one agent gets two items and the other gets nothing.
Now, consider allocation $X$.
Then for any item $g \in X_1$, we have 
\[\frac{v_2(X_2\cup\{g\})}{w_2} = 150 < \frac{v_2(X_1)}{w_1} = 200.\]
Hence, $X$ is not WWEF1.
For allocation $X'$, 
for any item $g \in X'_2$, we have 
\[\frac{v_1(X'_1\cup\{g\})}{w_1} = 200 < \frac{v_1(X'_2)}{w_2} = 600.\]
Hence, $X'$ is not WWEF1 either.
\end{proof}

\end{document}